\documentclass[reprint,prd,aps,twocolumn,superscriptaddress,eqsecnum,floatfix,preprintnumbers,amsmath,amssymb,
footinbib,longbibliography,citeautoscript]{revtex4-1}

\pdfoutput=1

\usepackage{graphicx}
\usepackage{epstopdf,epsfig}
\usepackage[scriptsize]{subfigure}
\usepackage[export]{adjustbox}
\usepackage{dcolumn}
\usepackage{bm}
\usepackage{marginnote}
\usepackage{changebar}
\usepackage{caption}
\usepackage[colorlinks]{hyperref}

\hypersetup{
    colorlinks=true,       
    linkcolor=blue,        
    citecolor=blue,        
    filecolor=black,       
    urlcolor=black,        
    bookmarks=false,
    pdffitwindow=true,
    pdfpagelayout=SinglePage,
    plainpages=false
}

\renewcommand{\d}{\mathrm{d}}

\newcommand{\be}{\begin{equation}}
\newcommand{\ee}{\end{equation}}

\newcommand{\ba}{\begin{equation}\begin{aligned}}
\newcommand{\ea}{\end{aligned}\end{equation}}

\newcommand{\bea}{\begin{eqnarray}}
\newcommand{\eea}{\end{eqnarray}}

\newcommand{\beg}{\begin{gather*}}
\newcommand{\eng}{\end{gather*}}

\newcommand{\angstrom}{\textup{\AA}}

\begin{document}

\title{Vacuum polarization at the tip of a water drop}

\author{Rouslan Krechetnikov}
\author{Andrei Zelnikov}
\affiliation{University of Alberta, Edmonton, Alberta, Canada T6G 2E1}

\date{\today}

\begin{abstract}
First rejected by the Nobel prize laureate Wolfgang Pauli as non-sense, the effect of vacuum polarization proves to be commonly observable in everyday life.
\end{abstract}

\maketitle

Back in 1961 the Soviet physicist Fedyakin discovered experimentally that water in small amounts possesses much higher viscosity than in ordinary state. After a decade of world-wide furor, in 1973 Derjaguin \& Churaev \cite{Derjaguin:1973} published a disclaimer attributing the anomalous ``polywater'' properties to impurities. At the time, Feynman remarked that should such a material be present, then ``an animal would exist that ingests water and excretes polywater, using the energy released from the process to survive''. A similar idea was exploited in Vonnegut's 1963 novel Cat's Cradle, in which the fictional ice-nine was a form of water that solidified any liquid water it contacted at room temperature, thus making it capable to destroy all life on Earth. Around the same time, in 1971 a seemingly unrelated theoretical discovery was made by Huh \& Scriven \cite{Huh:1971} that there is an unexplained singularity of an infinite friction force near the moving contact line of a rain drop sliding down the glass window, which they aphoristically paraphrased as ``not even Herakles could sink a solid''. Since then, while still maintaining the idea that water should behave ordinarily near the contact line, enormous experimental and theoretical efforts were put to relax the singularity, primarily with the help of introduction of a slip velocity near the contact line or a pre-cursor film. In our recent work, with the quantum field theory we demonstrated that near the contact line, due to increasingly dominant effects of vacuum polarization, the quantum fluctuations make water behave very differently from its ordinary state and, in fact, act more like a solid in resurrection of the polywater-like notion. As a result, by abandoning, now at the rigorous basis, the idea that water is in its normal state near the contact line, the Huh-Scriven singularity of the friction force is naturally resolved at the microscopic level.

\section{Huh-Scriven paradox}

\begin{figure}
\centering
\includegraphics[height=0.85in]{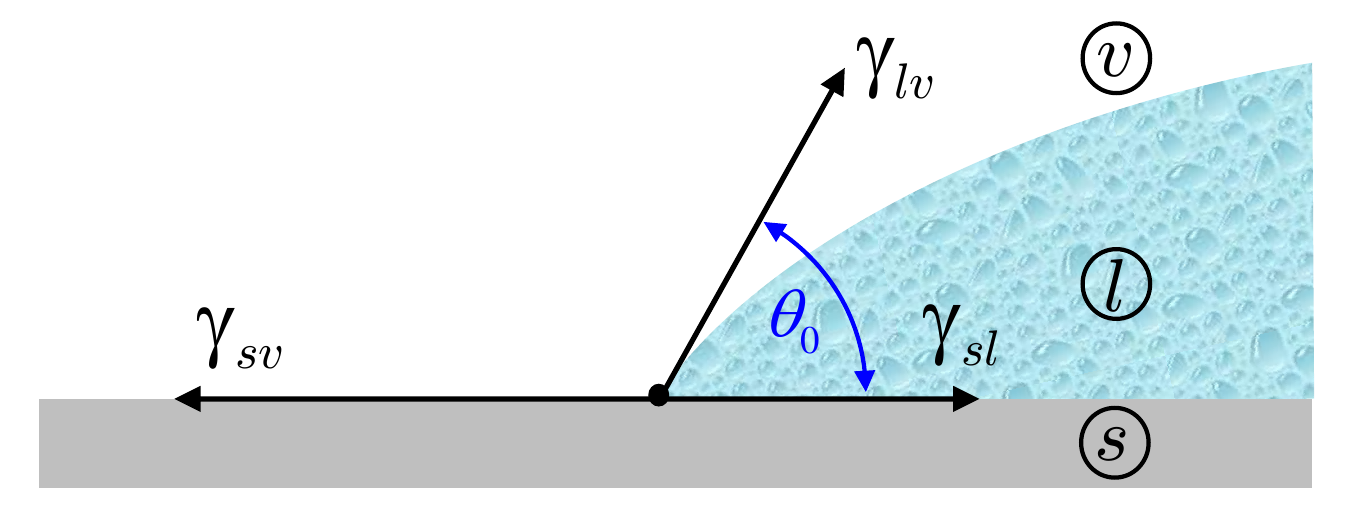}
\caption{Young's diagram at the contact line of a liquid drop.}\label{fig:Young-diagram}
\end{figure}

When a drop of viscosity $\mu$ moves down a glass window with speed $U$, the shear stresses scale with the distance $r$ to the contact line as $\sim \mu \, U / r$ and hence formally diverge at the tip of the drop, cf. figure~\ref{fig:Young-diagram}. The rate of energy dissipation $\dot{E}$ per unit contact line length is then
\begin{align}
\d \dot{E} \sim \mu \, U^{2} \frac{\d r}{r} = \mu \, U^{2} \d \ln{r},
\end{align}
and hence diverges logarithmically when integrated in the neighborhood of the contact line. This constitutes the famous Huh-Scriven paradox \cite{Huh:1971}.
The wonders about the contact line began since Young's work \cite{Young:1805} in 1805, who introduced the force diagram $\gamma_{sl}-\gamma_{sv} + \gamma_{lv} \cos{\theta_{0}} = 0$ for the solid-liquid-vapor point, leading to the puzzle as to where do these forces actually apply. The fact that there are three surface tensions pulling in different directions contradicts the fact that liquid at rest must be isotropic according to Pascal's principle. Anisotropy implies that liquid near the triple point must behave as a solid capable of maintaining shear stresses at rest. Hence, it becomes clear that this small, but finite, volume of matter acts as an anchor for Young's force diagram. But how is it possible that liquid may behave as a solid at room temperature?

\vspace{-0.15in}

\section{Origin of surface tension} \label{sec:ST}

\vspace{-0.1in}

\begin{figure}
\centering
\includegraphics[height=1.25in]{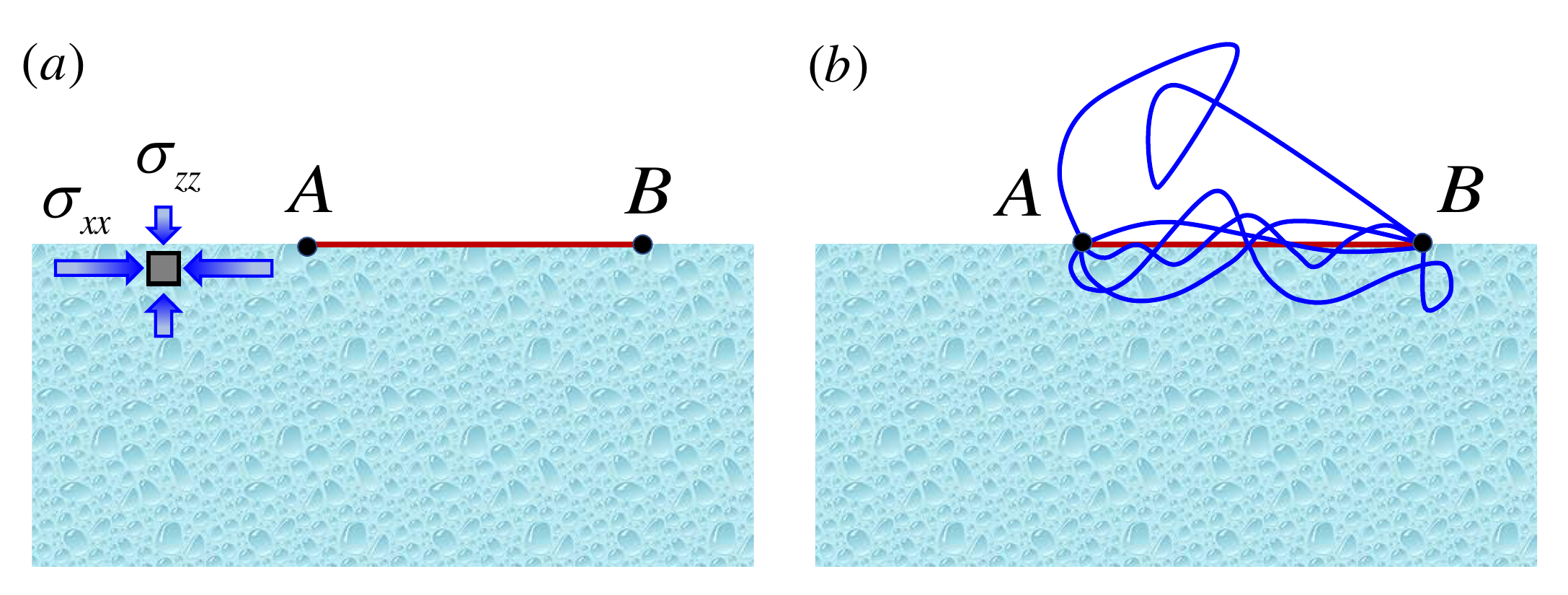}
\caption{\setlength{\rightskip}{0pt} Feynman path integral replaces the classical unique trajectory (a) for a system with a functional integral over an infinity of quantum-mechanically possible trajectories (b).}\label{fig:path-integral}
\end{figure}

Surface tension $\gamma_{lv}$ of a liquid results from the integral difference between normal $\sigma_{zz}$ and tangent $\sigma_{xx}$ to the interface ($z=0$) quantum electromagnetic stresses resulting from van der Waals attraction, cf. figure~\ref{fig:path-integral}(a),
\begin{align}
\label{ST:calculation}
\gamma_{lv}=\int_{-\infty}^{\infty}{\left[\sigma_{xx}(z)-\sigma_{zz}(z)\right] \, \d z}.
\end{align}
Obviously, should stresses in the liquid be isotropic, $\sigma_{xx}=\sigma_{zz}$, there would be no surface tension. The stresses $\sigma_{xx}$ and $\sigma_{zz}$ are computed with the Lifshitz theory as an extension of Casimir's work to real materials.

Accordingly, the contribution of quantum mechanics goes beyond the obvious fact that quantum physics underlies the effective intermolecular potentials (e.g. Lennard-Jones) employed in the classical explanations of surface tension. Namely, equation \eqref{ST:calculation} indicates that even if the interface is between a dielectric and the vacuum, polarization of the latter also contributes to surface tension -- this is not an artefact of the theory as the measurable Casimir effect \cite{Casimir:1948b} is also due to the vacuum polarization and zero-point energy of quantum fluctuations. In particular, quantum field theory predicts that gravitational effects of the zero-point energy are crucial at early stages of evolution of our Universe and may be relevant to the dark energy problem as well. In 1988 the future Nobel prize laureate Kip Thorne suggested that Casimir energy could be used to stabilize a wormhole and thus lead to the possibility of interstellar travel. In Kaluza-Klein theory, Casimir effect provides a mechanism for spontaneous compactification of extra spatial dimensions.

To understand the origin of the vacuum polarization near the interface, recall that liquids consist of molecules interacting via exchange of photons. At subcritical temperatures, these photons are primarily virtual, i.e. not on mass-shell and, thanks to the Heisenberg uncertainty principle, for a very short time capable of borrowing necessary energy from vacuum in order to propagate to the other molecule and return the borrowed energy back to the vacuum. In the Feynman path integral formulation of quantum mechanics, cf. figure~\ref{fig:path-integral}, two molecules $A$ and $B$ interact by exchanging virtual particles not only along the shortest path from A to B (a), but also along a bunch of all possible trajectories, including those passing through vacuum (b). Although virtual particles are \textit{unobservable}, they feel the presence of matter and inhomogeneity of space, such as an interface between matter and vacuum, and their cumulative action leads to \textit{observable} effects, in particular, to that of Casimir and the quantum vacuum polarization outside the real material!

The latter proves to be crucial for accurate computation of surface tension, but required several decades to develop mathematical methods, specifically spectral geometry \footnote{{b}y DeWitt, who was startled with a seminar given by Casimir around 1960, in particular that zero-point energy gives the force between metal plates, which he used to think was unphysical.}, in order to properly extract from the Lifshitz theory ultra-violet renormalizable divergencies associated with the isotropic zero-point energy of the vacuum as well as non-renormalizable ones, which account for the unresolved physics at the scales shorter than interatomic distances $r_{\mathrm{m}}$ and are responsible for the anisotropic surface tension phenomena \cite{Zelnikov:2021}. Given different scalings with the interfacial thickness $w$ of subleading $\sim (w \, r_{\mathrm{m}})^{-1}$ and finite $\sim w^{-2}$ contributions to surface tension, our theory is testable near the critical point and shows reasonably good agreement with experimental data. Also, while in 1978 Schwinger et al. \cite{Schwinger:1978} reported the calculated magnitude of surface tension for liquid helium three times higher than the actual one $\gamma_{lv}=0.37 \, \mathrm{mN \, m^{-1}}$, our calculations produce $0.48 \, \mathrm{mN \, m^{-1}}$ based on the known interface thickness $6.5 \, \angstrom$ and interatomic distance $3.75 \, \angstrom$. From a practical point of view, accurate computation of forces and stresses near interfaces is instrumental in developing micro- and nanomechanical systems, in particular.

\vspace{-0.15in}

\section{How to sink a solid}

\vspace{-0.1in}

As envisioned in \S \ref{sec:ST}, application of the Lifshitz theory to the contact line problem in figure~\ref{fig:Young-diagram} indeed shows that the stress distribution is no longer isotropic \cite{Krechetnikov:2022}. Also, for most common solid-liquid-vapor triple-point combinations, the microscopic contact angle either collapses to $0$ or unfolds to $\pi$, which corresponds to perfect wetting and nonwetting, respectively. The macroscopic contact angle $\theta_{0}$ in Young's equation is different from the actual one, at which the interface meets the substrate, and instead is set asymptotically at the distances $r$ where quantum effects become negligible. Our theory reveals the quantum nature of the forces governing the contact line behavior, being the result of dominant vacuum polarization. One can say that compared to the fluid in the bulk, fluid's behavior near the wedge tip is anomalous due to anisotropy of the stresses thus resurrecting Fedyakin's suggestion that water in sufficiently small amounts behaves abnormally.

Returning to the Huh-Scriven paradox in the case of a wedge dynamically moving with velocity $U$ along the substrate, the viscous stresses $\sim \mu \, U / r$ are on the order of the quantum electromagnetic stresses at $r^{*} \sim \left(A_{\mathrm{H}} / 6 \pi \mu U\right)^{1/2} \sim 100 \, \mathrm{nm}$, where $A_{\mathrm{H}}$ is the Hamaker constant taken for water on mica. Below this scale, the quantum stresses dominate and enable tearing the ``solidified'' liquid off the substrate when the contact line is receding and bulldozing when it is advancing, thus naturally resolving, at the microscopic level, the Huh-Scriven paradox of a singular friction force. It transpires that the paradox arose from the assumption that liquids behave ordinarily regardless of the distance $r$ to the contact line.

\vspace{-0.15in}

\section{Vacuum polarization around us}

\vspace{-0.1in}

In our everyday life we are used to ignoring the fact that our Universe is quantum, because it seems to be much more comfortable to think that we live in a deterministic classical world. But only ``on the surface'' nature appears classical. Above discussion suggests that one can literally touch quantum mechanics on surface of liquids because of the quantum origin of surface tension. While quantum mechanics is usually associated with microscopic systems, its applicability happens to go well beyond small scales. For example, the Chandrasekhar limit -- the star mass above which electron degeneracy pressure in star's core becomes insufficient to balance its own gravitational self-attraction -- is of quantum origin due to Pauli's exclusion and Heisenberg's uncertainty principles. On even larger scales, the galaxies were formed from primordial quantum fluctuations that were amplified during the early Universe expansion. The vacuum is not empty but instead ``alive'', i.e. made of quantum fluctuations with particles popping into and out, and even our own existence is the consequence of that.



%

\end{document}